\documentstyle{aipproc}
\input epsf       

\begin{document}

\centerline{{\it Proceedings of the 5$^{th}$ Huntsville Gamma-Ray Bursts Symposium (1999)}}
\vspace{0.1cm}

\title{A Robust Filter for the BeppoSAX Gamma Ray Burst Monitor Triggers}

\author{M.~Feroci$^{(1)}$ C.L.~Bianco$^{(1)}$ F.~Lazzarotto$^{(1)}$ A.~Mattei$^{(1)}$ G.~Ventura$^{(1)}$ E.~Costa$^{(1)}$ and F.~Frontera$^{(2),(3)}$}
\address{$^{(1)}$Istituto di Astrofisica Spaziale - CNR, Via del Fosso del Cavaliere, Roma, Italy\\
$^{(2)}$Istituto Tecnologie E Studio Radiazioni Estraterrestri - CNR, Via Gobetti 101, Bologna, Italy\\
$^{(3)}$Dipartimento di Fisica, Universit\'a di Ferrara, Via del Paradiso 12, Ferrara, Italy}

\maketitle

\begin{abstract}
The BeppoSAX Gamma Ray Burst Monitor (GRBM) is triggered any time a 
statistically significant counting excess is simultaneously revealed by at
least two of its four independent detectors.
Several spurious effects, including highly ionizing particles crossing
two detectors, are recorded as onboard triggers. In fact, a large number
of false triggers is detected, in the order of 10/day.
A software code, based on an heuristic algorithm, was written to
discriminate between real and false triggers. We present the results
of the analysis on an homogeneous sample of GRBM triggers, thus
providing an estimate of the efficiency of the GRB detection 
system consisting of the GRBM and the software.

\end{abstract}

\section*{The Gamma Ray Burst Monitor}

The GRBM \cite{frontera97,amati97,feroci97,costa98}
is a gamma-ray detection system onboard the
BeppoSAX satellite. 
It is the secondary
function of the anticoincidence shields of the PDS experiment. It
is composed of four $\sim$1100 cm$^{2}$ slabs of CsI(Na) scintillators
operating in the 40-700~keV range. Each detector provides time
series of the detected counts in the above energy range, with 1 s
resolution as a continuous housekeeping and with better than
8~ms resolution upon trigger. An oboard trigger is active whenever a
statistically significant counting excess is simultaneously detected in
at least two of the four detectors.

\section*{Aims, Method and Limits}

Besides of being gamma-ray detection devices,
the scintillating crystals composing the GRBM are sensitive
to highly ionizing particles that, leaving a large amount of energy
($\sim$GeV) in just one shot, result in a phosphorescence phenomenon,
with a consequent detection by the electronics of a large number
of counts in few tens of ms, i.e. a {\it spike} in the counting rate.
When the same particle subsequently crosses two detectors, it
causes the onboard logic to trigger on the event, that is
electronically undistinguishable from a cosmic gamma-ray
transient (no particle anticoincidence is available to the GRBM).
The number of spikes that trigger the onboard electronics
(originating false triggers) is of the order of 9-10/day.
Therefore, the aim of this work is to provide a software filter
that allows to make a `safe' first-order discrimination
of the `instrumental' triggers from the `cosmic-origin' triggers.
This will allow the reduction of the huge number of onboard triggers
recorded so far (of the order of 10,000 in the first three years
of BeppoSAX operations), and to apply a more refined program
to the generated sample of triggers. 
A first drawback of the `roughness' of our filter is that it does
not include criteria to separate gamma ray bursts from solar 
flares and from soft gamma repeaters events. This task will
be carried out by the `second order' filter.

Our software filter is based on the
automatic on-ground analysis of the high resolution time series,
according to criteria established on the basis of the known
detector/electronics behaviour and an extended study of the
GRBM time series.
Usually, an eye inspection of a GRBM light curve is sufficient to
discriminate cosmic gamma-ray events from spurious events. 
However,
when an archival search for real GRB is carried out this becomes
not viable anymore, and an automatic filter is needed.
We therefore set-up an IDL-based software code that 
implements a number of discrimination criteria to the 
GRBM light curves. This is a first approach to the problem
and we did not make use of all the information that 
the GRBM provides for each onboard trigger. In fact,
also 1-s resolution data are available in two energy ranges
(40-700 and $>$700 keV) and time-averaged energy spectra
but they are not used in the analysis presented here.

The criteria implemented in the program are based on the
knowledge of the instrument operational principles and on 
experience on the observed light curves. 
It is likely that particular cases exist that
were not taken into account in our code. 
At this time, the following parameters of the individual events 
are computed for each of the four GRBM detection units: 
duration, rise time,
simultaneity, shape, full width at half maximum.
The basic criterium of the code is to assign a `score' to 
each of these parameters, based on their comparison
between pairs of detectors,
with the goal of having the smallest
score to the most-likely-cosmic events. The individual scores for 
the different parameters then combine together to give a total score
accounting for all the measured event characteristics.
For the final score, the higher is the value the smaller
is the probability of being a cosmic event. 

\section*{Calibration of the filter}

In order to test and calibrate the efficiency of our software
filter, we set-up a sample of selected events whose origin
was known by different methods (typically BATSE or IPN events).
To these a number of eye-screened false triggers 
were added. 
In Figure 1 we present the results of the analysis carried out
with our filter on such a sample of events. 
The x-axis reports the score given to each event by the
filter. The texture classifies the events, based on the comparison
with data from other satellites (BATSE, Ulysses, etc.).
The sample includes several short GRBs in order to test the
ability of the software to distinguish between them and spikes, but
many long GRBs are also included. Thus,
we can define 3 score classes (Burst, Spike,
Doubt), with the confidences given in Table 1 (the sum of
every row is 100\%), computed assuming multinomial distribution.
The software works on single peaks even during the
evolution within a multiple-peaked event. Therefore, the results
reported in the Table should be intended as applicable to
individual peaks in each light curve.
Examples of events belonging to these three categories 
are presented in Figure 2.

\begin{table}
\begin{tabular}{cccc}
Score           & Burst     &  Spike   &  Doubt \\
\tableline
Score $<$2        & $(83.3\pm5.8) \%$ & $(9.6\pm4.5) \%$ & $(7.1\pm4.0) \%$  \\
2$\leq$Score$<$4  & $<$4.5\% (conf.lev. 68\%)  & $(16.0\pm7.3) \%$ & $(84.0\pm7.3) \%$  \\
Score $\geq$4     & $(1.7\pm1.0) \%$ & $(94.0\pm1.8) \%$ & $(4.2\pm1.6) \%$  \\
\end{tabular}
\caption{Score calibration for the software filter}
\label{table1}
\end{table}

\begin{figure}
\epsfxsize=14.0cm \centerline{\epsfbox{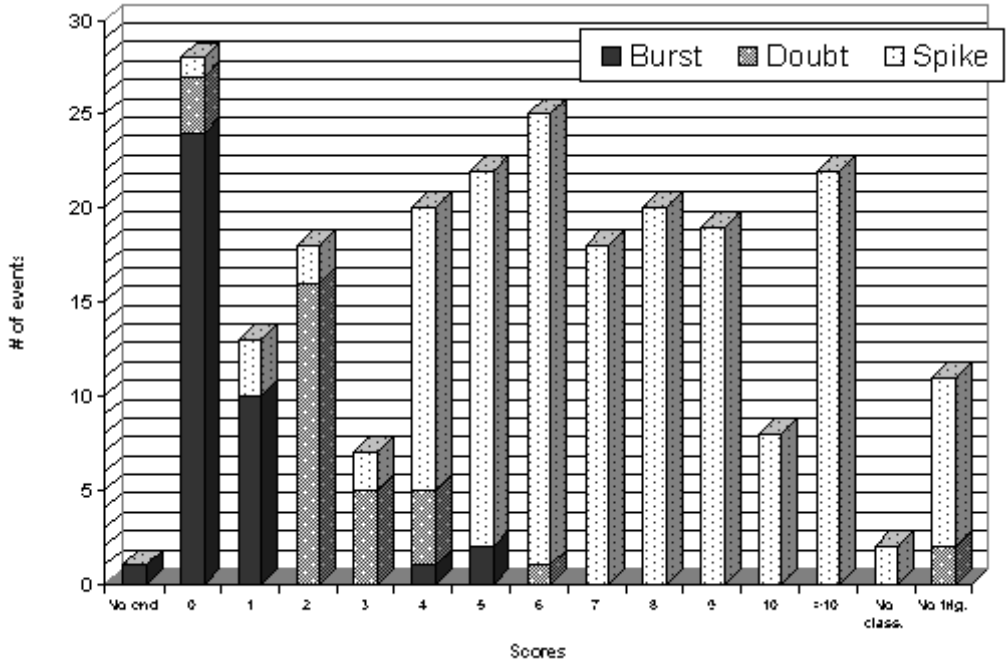}}
\caption{Efficiency diagram of the software filter: 
probability of true cosmic events is inversely proportional
to the score value on x-axis}
\label{efficiency}
\end{figure}

There are 3 additional special classes in the plot: 
No-End, No-Class, No-Trig.
No-End: the signal does not return to noise level
before the end of the light curve (GRBM high time resolution
data have a maximum coverage of 106 s). This can be due to
long bursts, 
or to a variable background (see an example in \cite{feroci97}).
No-Class: the signal has not been analyzed. This
can happen for weak events if the software is not able to estimate 
the duration of the signal.
No-Trig: the software is unable to find other than
noise in the light curve, possibly due to very weak signals.

\begin{figure}
\epsfxsize=8.4cm \centerline{\epsfbox{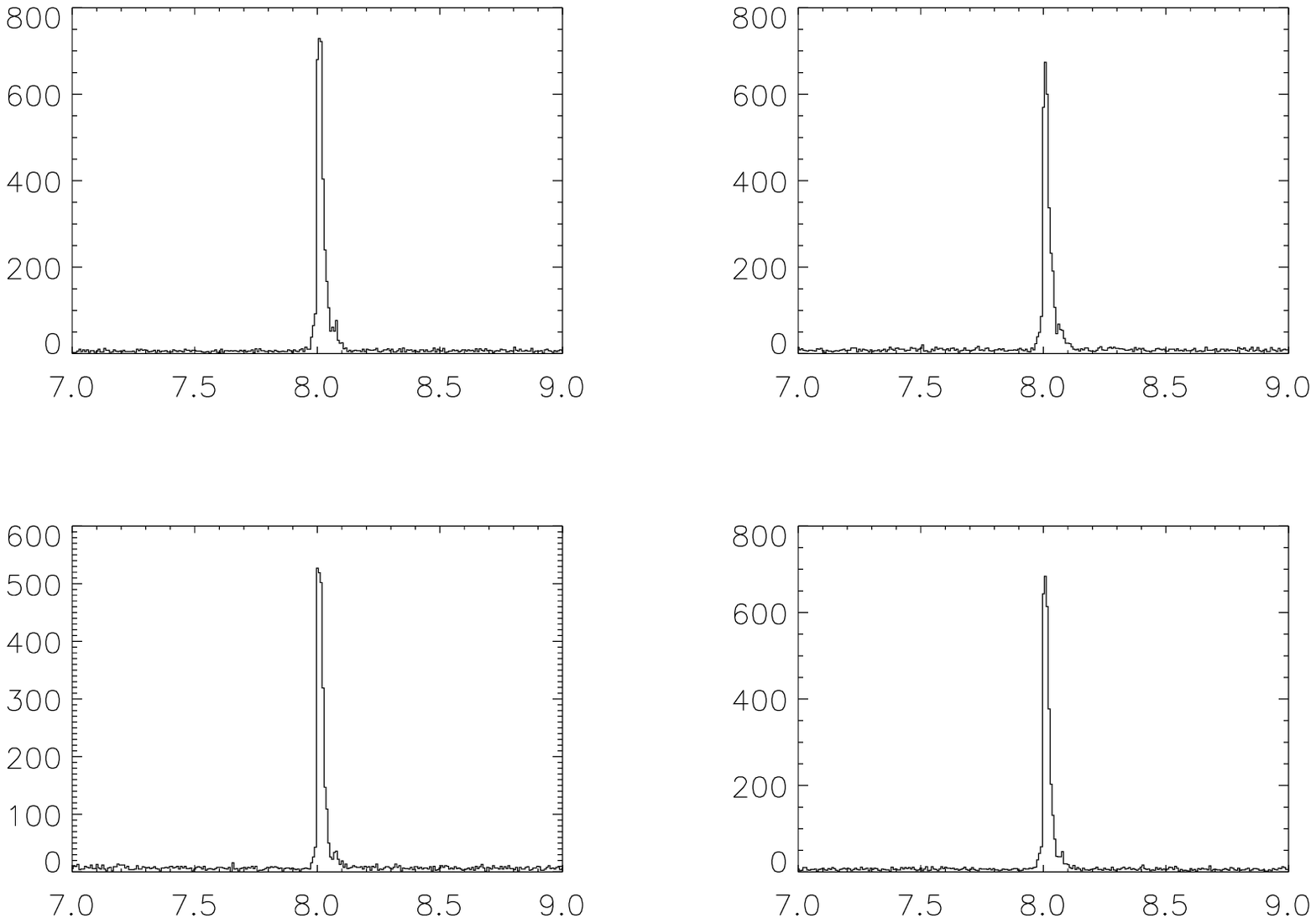}}
\epsfxsize=8.4cm \centerline{\epsfbox{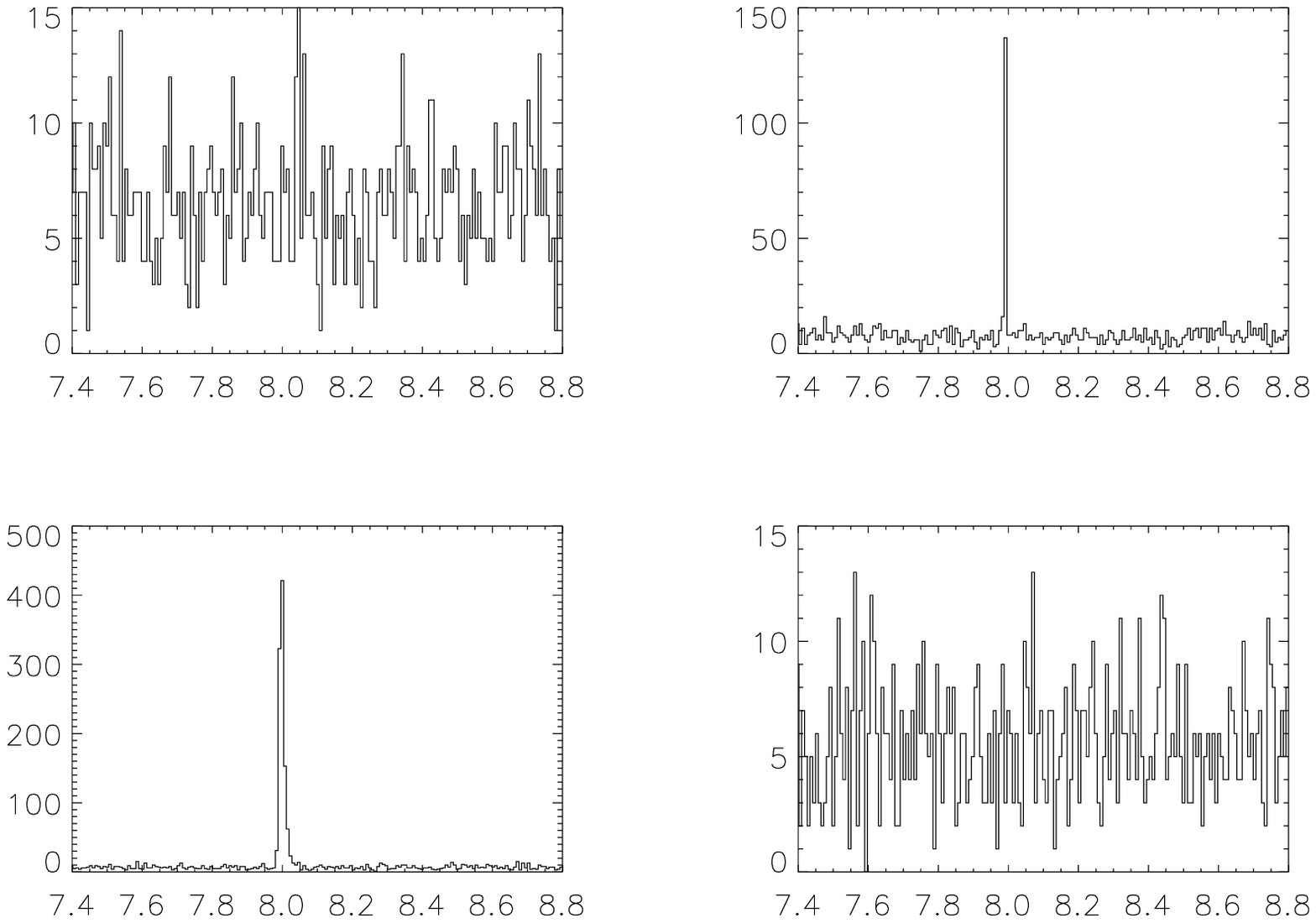}}
\epsfxsize=8.4cm \centerline{\epsfbox{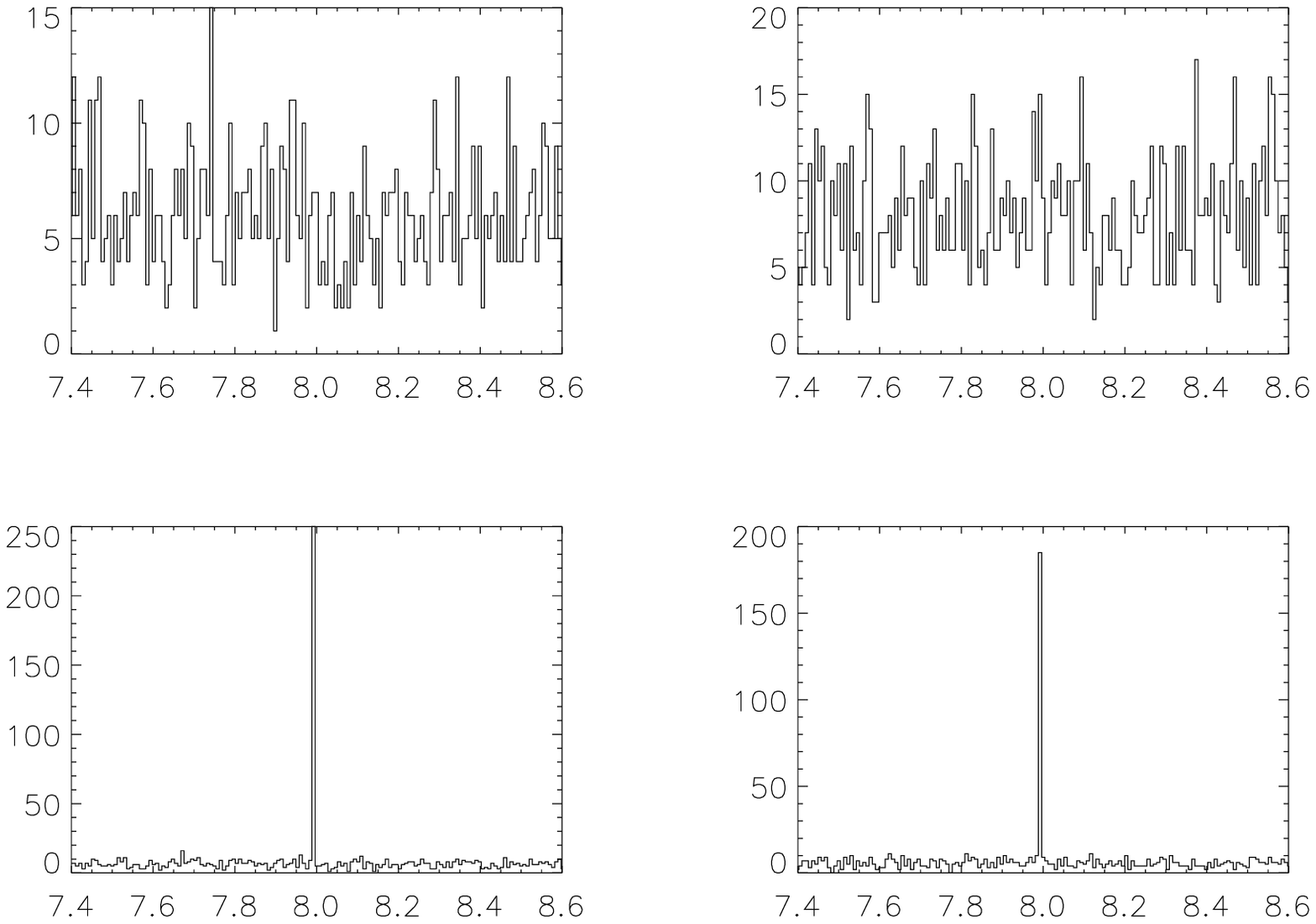}}
\caption{
8-ms resolution light curves in the four GRBM detectors, for
a real short gamma ray burst (Top 4 panels), a spike (Middle 4 panels) and
a dubious event (Bottom 4 panels). X-axis is time in seconds, Y-axis
gives the counts/bin.}
\label{real}
\end{figure}

\section*{Application to the GRBM database}

The software filter has been applied to a fraction of the
BeppoSAX GRBM data archive, covering more than 3 years of
elapsed time, but consisting of about 603 days of satellite
observing time. The goal was to test our software on an
homogenous sample of onboard triggers. A noticeable by-product
of this operation is an estimate of the GRB detection efficiency. 
As stated above, the filter operates on individual
peaks. The result of the analysis gives: 440 peaks from cosmic events,
510 doubt cases, 4648 spikes, 58 No End, 86 No Class.
If we apply the filter efficiency
that was defined in the previous section, then the number of
cosmic peaks become (367$\pm$26). They belong to about 340
individual events, of which only $\sim$180 have been post-facto 
verified to be real cosmic events.

\section*{Summary and Conclusions}

The results of the analysis on an unselected portion ($\sim$40\%) of the
BeppoSAX/GRBM data archive presented in this paper allow to
draw the following conclusions:

- A number of $\sim$180 cosmic events detected by the
BeppoSAX/GRBM were identified (including a small number
of Solar Flares and events from Soft Gamma Repeaters),
over a net exposure time of $\sim$545 days, 
leading to an estimation of the {\it GRBM
efficiency on triggering cosmic events of $\sim$0.33/day}
(additional events are detected, but without an onboard trigger).

- The automatic filter has a $\sim$50\% efficiency
on selecting real events out of false triggers (i.e., any event selected
by the program has a $\sim$50\% probability of being of cosmic origin). 
This reduced efficiency is likely due to the incompleteness of the sample 
on which the code was calibrated.

- The automatic filter has a $>$90\% efficiency
on descarding false triggers (i.e., any event not selected by the 
program has a less than 10\% probability of being a real event,
based on the software calibration). 
So, the filter allows to reduce the number of light curves 
to be analyzed to about 
10\% of the total, with an expected efficiency of more than 90\%.


\begin{references}

\bibitem{frontera97}Frontera~F. et. al. A\&A~Suppl.~Ser., {\bf 122}, 357, 1997

\bibitem{feroci97}Feroci~M., et. al. Proc. SPIE, {\bf 3114}, 186, 1997

\bibitem{amati97}Amati~L., et. al. Proc. SPIE, {\bf 3114}, 176, 1997

\bibitem{costa98}Costa~E., et. al. Adv. Space Res. {\bf 22}, 1129, 1998


\end{references}
\end{document}